\documentclass[epj,referee]{svjour}
\usepackage{graphics}
\begin{document}

\title{On the rich--club effect in dense and weighted networks}
\author{
V. Zlatic\inst{1,2}
\and G. Bianconi\inst{3}
\and A. D\'{\i}az-Guilera\inst{4}
\and D. Garlaschelli\inst{5}
\and F. Rao\inst{6,7}
\and G. Caldarelli\inst{1,4,8}
}
\institute{
INFM-CNR Centro SMC and Dipartimento di Fisica, Universit\`a di Roma La Sapienza, Piazzale Moro 5, 00185 Roma, Italy.
\and
Theoretical Physics Division, Rudjer Bo\v{s}kovi\'{c} Institute, P.O.Box 180, HR-10002 Zagreb, Croatia.
\and
The Abdus Salam International Centre for Theoretical Physics, Strada Costiera 11 34014 Trieste, Italy.
\and
Dept. de F\'{\i}sica Fonamental, Facultat de F\'{\i}sica, Universitat de Barcelona Diagonal 647 08028 Barcelona, Spain.
\and
Dipartimento di Fisica, Universit\`a di Siena, Via Roma 56, 53100 Siena, Italy.
\and
Centro Studi e Museo della Fisica Enrico Fermi, Compendio Viminale, 
00185 Roma, Italy.
\and
Universit\'e ``Louis Pasteur'' Laboratoire de Chimie Biophysique/ISIS
8, allee Gaspard Monge - 67000 Strasbourg, France.
\and
Linkalab, Center for the Study of Complex Networks, 09100 Cagliari, 
Sardegna, Italy.
} 
\date{Received: date / Revised version: date}
%
\abstract{
For many complex networks present in nature only a single instance, usually of large size, is available. Any measurement made on this single instance cannot be repeated on 
different realizations. In order to detect significant patterns in a real--world network it is therefore crucial to compare the measured results with a null model counterpart. Here we focus on dense and weighted networks, proposing a suitable null model and studying the behaviour of the degree correlations as measured by the rich-club coefficient. Our method solves an existing problem with the randomization of dense unweighted graphs, and at the same time represents a generalization of the rich--club coefficient to weighted networks which is complementary to other recently proposed ones.
\PACS{
      {89.75.Hc}{Networks and genealogical trees}   \and
      {89.75.Fb}{Structures and organization in complex systems}
     } 
} 
\maketitle
\section{Introduction}
Networks, identified by the set of connections (edges, or links) between units (vertices) of a system, are widespread in nature  \cite{Book,siam}. One 
of the most important local quantities is the number of connections of vertex $i$, called the degree $k_i$.
Real complex networks are characterised by specific features such as small-world 
effect  \cite{SW} and scale-invariant degree distribution \cite{RMP}. 
At the same time for many networks (some maps of the Internet \cite{CMP01}, 
or the whole WWW \cite{AJB01} or biological webs \cite{GCP03})
additional information on the intensity of connections is available. 
In this case, one can assign a weight $w_{ij}$ to the link between vertices $i$ and $j$. The weighted counterpart of the degree is given by the strength $s_i=\sum_j w_{ij}$.

\begin{figure}[b]
\centerline{
\resizebox{0.9\columnwidth}{!}{%
  \includegraphics{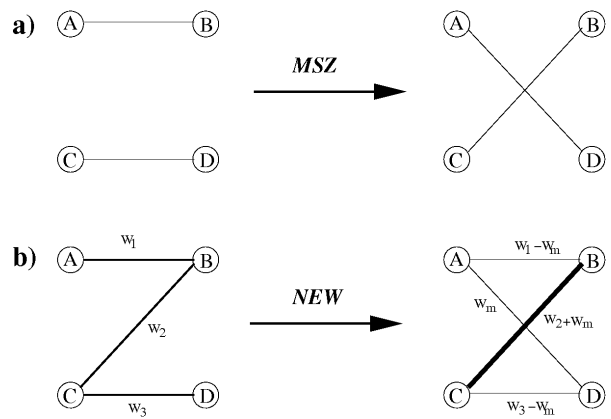}
}
}
\caption{({\bf a}) The rewiring in the MSZ algorithm. Note that no other links are 
present between the vertices. ({\bf b}) The rewiring in our new procedure transfers the weight also between existing links. In unweighted networks this implies the creation of multiple links. Note that if $w_1=w_m$  this would correspond to the removal of the link between $A$ and $B$.}
\label{fig1}
\end{figure}

An important quantity, which is the main focus of the present work, is the 
{\em rich-club coefficient} (RCC) $\phi(k)$  introduced as 
a correlation measure of the interconnectivity between nodes with a 
``large'' degree \cite{zu}. 
In particular, given the number $E_{>k}$ of edges 
between the $n_{>k}$ vertices whose degree is larger than $k$ the RCC is defined as
\begin{equation}
\phi(k)\equiv \frac{2 E_{>k}}{n_{>k}(n_{>k} -1)}.
\end{equation}
In other words the RCC $\phi(k)$ measures the 
probability that the edges between pairs of vertices whose degree
is larger than $k$ are actually drawn. 
If no edge is present then $\phi(k) = 0$, while if all the possible edges are present then $\phi(k) = 1$. 
The interest for this quantity comes from the fact that an increasing trend for the RCC would reveal the presence of correlations 
between large hubs. However it is easy to check that, even without any correlation, 
it is more likely that one edge is shared between two hubs  rather than 
between two vertices with a low degree.
In particular one finds that in an uncorrelated graph the RCC increases as $k^2$ for large $k$ \cite{col}.
To overcome this problem, Colizza et al.  \cite{col} considered a suitable set of randomized versions of a given graph, and computed the corresponding RCC $\phi^{rand}(k)$ on this ensemble. They proposed to consider the ratio 
\begin{equation}
\rho(k)\equiv\frac{\phi(k)}{\phi^{rand}(k)}
\end{equation} 
in order to assess the presence of correlations in a variety of systems.
The randomizing strategy adopted preserves the degree sequence of the original graph 
by using an algorithm proposed by Maslov, Sneppen and 
Zaliznyak (MSZ)  \cite{science,MSZ04}.  
Two edges are selected, and one endvertex of the first is exchanged with one endvertex of the second and viceversa, as shown in Fig.\ref{fig1}{\bf a}. 
This local rewiring is successful only if no edge already exists in place of the rewired ones.
Otherwise, two new edges are selected and a new rewiring is attempted.

This rewiring procedure presents some limitations in the case of dense graphs. 
Consider the limit case of a complete graph with $N$ vertices:  
here no rewiring is possible 
and the randomized set of graphs corresponds to the single original instance. 
Therefore one always finds a value $\rho(k)=1$ for every $k$, even if only the single value $k=N-1$ is present in the network.  
However, this merely reflects the fact that the configuration space of the possible randomizations is reduced 
to only one configuration. 
For dense graphs, such as the World Trade Web \cite{originalWTW,WTW2,chatta} we consider later on and some  food webs \cite{GCP03}, the number of available configuration is also small. 
Furthermore, note that most collaboration networks  \cite{siam} (co--authorship, board of directors, movie actors) are obtained as projections of a bipartite graph.
For example in co--authorship networks there are two classes of vertices: the papers and the scientists authoring them. Links only exist between vertices belonging to different classes, and the graph is said to be bipartite.
From this bipartite graph one can obtain a one--mode network whose vertices represent only the scientists,  
and all scientists co--authoring at least one paper are connected in a clique (fully connected subgraph). Thus, even if the network is on average sparse, it is divided into dense cliques. 
In all these situations, any pair of edges belonging to the most important subunits (the cliques) cannot be rewired, and the ratio $\rho^{MSZ}(k)=\rho(k)=\frac{\phi(k)}{\phi^{MSZ}(k)}$ remains close to $1$ and cannot be used to assess the significance of the results. 

\section{New randomization and definition for the rich-club coefficient}
To overcome this problem we propose to generalize the MSZ algorithm and the definition of 
the rich--club coefficient by explicitly transforming 
the graphs analysed into weighted graphs.
Our proposal is then to consider the network as weighted, and redistribute the weights of the edges (possibly also on existing links) preserving the strength of the vertices as follows. 
Specifically, if $w_m$ is the minimum value of the weights in the original network 
we can modify the rewiring procedure by changing by an amount $w_m$ the weights of two pairs 
of edges  as shown in Fig.\ref{fig1}{\bf b}. 
This randomization (denoted hereafter as NEW) corresponds to the 
original MSZ algorithm when the network 
is sparse and unweighted. Indeed in this case the minimum weight $w_m$ is one and 
it is unlikely to draw twice the same edge.
We then make a second generalization by defining a rich-club coefficient for 
the weighted case as 
\begin{equation}\label{eq_phiw}
\phi_w(s)\equiv\frac{2 E_{>s}}{n_{>s}(n_{>s} -1)}.
\end{equation}
where $E_{>s}$ is the number of weighted edges between vertices whose {\em strength} is 
larger than $s$ and $n_{>s}$ is the number of vertices whose 
{\em strength} is larger than $s$.
We finally define 
\begin{equation}
\rho^w(s)\equiv\frac{\phi_{w}(s)}{\phi_w^{NEW}(s)}
\end{equation}
We stress that our focus here is different from other recent generalizations of the rich--club coefficient, which are specific to weighted networks  \cite{serrano2,colizza2}.
Our approach originates from the aforementioned problem with the randomization of dense unweighted networks. 
The above definition, when coupled with our new randomization procedure, solves this problem even for complete graphs. Moreover, it
automatically recovers the original (and intuitive) MSZ approach for sparse unweighted networks.
Additionally, it provides a coherent generalization of the rich--club to weighted networks, emphasising patterns which are complementary to those investigated by other proposals  \cite{serrano2,colizza2}. 
In our above definition we are considering only the {\em numbers} $E_{>s}$ of edges and $n_{>s}$ of vertices, and not their weight. 
This captures a structural property of many social and/or 
economic networks where only the presence of a tie, even if weak \cite{grano83}, plays a role in the organization of the system. If combined with the other available definitions, this property allows to recover a richer picture of the correlation structure of weighted networks, as we show below for a particular example. Finally, the choice of the minimum value $w_m$ automatically sets a scale for the weights in the randomized ensemble. In this way we preserve the original strengths without the need to regard each weighted edge as a superposition of binary edges and split them as in other approaches  \cite{serrano2}. Indeed, the latter procedure is very sensitive to the (arbitrary) choice of the unit of weight  \cite{newman_multiple}.
Remarkably, through a single definition we can address all the above problems and unambiguously characterize any network, ranging from unweighted to weighted, and from sparse to dense.

Note that in our randomization strong links present in the original network can be diluted into more and weaker links in the randomized ensemble. However, the strength sequence $\{s_i\}_{i=1}^N$, representing the vector of the strengths of all vertices, is preserved at any instance. As a result, it is possible to exploit the results for randomized networks with preserved strength sequence \cite{bosonic} to characterize how a particular strength distribution affects the properties of the randomized networks. The main lesson is that weighted properties, including the weighted clustering coefficient, the weighted average nearest neighbour degree and the disparity, display a kind of Bosonic behaviour that parallels the Fermionic one observed for unweighted networks with preserved degree \cite{bosonic}. 
In the present study we are interested in understanding how the rich-club coefficient of the original network relates to that of the randomized variants. 

Also note that, among the other randomization discussed in the literature, one in particular showed that preserving both the degree sequence and the modularity is in most cases enough to explain the correlations of the network \cite{guimera}.
Due to the close relation between degree correlations and the rich-club coefficient, we expect that such a randomization would preserve the rich-club as well. 
Our perspective is however in some sense reversed: we shall regard dense subgraphs as arising in a bottom-up manner from the local constraints, rather than as forcing the rich-club to emerge in a top-down fashion from the modularity.
In particular, we expect initially dense (either topologically or in a weighted sense) modules to dilute, and a possible reordering of vertices among modules for each randomized variant.

In the same spirit, our method is different from randomizations that step back to the bipartite structure of collaboration or affiliation networks, rewire links preserving the degrees in the bipartite graph, and then project the latter to obtain a randomized one-mode network. In both methods, cliques that are present in the original network will not be preserved. However, the bipartite randomization will create other cliques in any variant of the network, arising as the two-mode network is projected onto the one-mode one. 

\section{Sparse and dense unweighted graphs}
\begin{figure}[t]
\centerline{
\resizebox{0.9\columnwidth}{!}{%
  \includegraphics{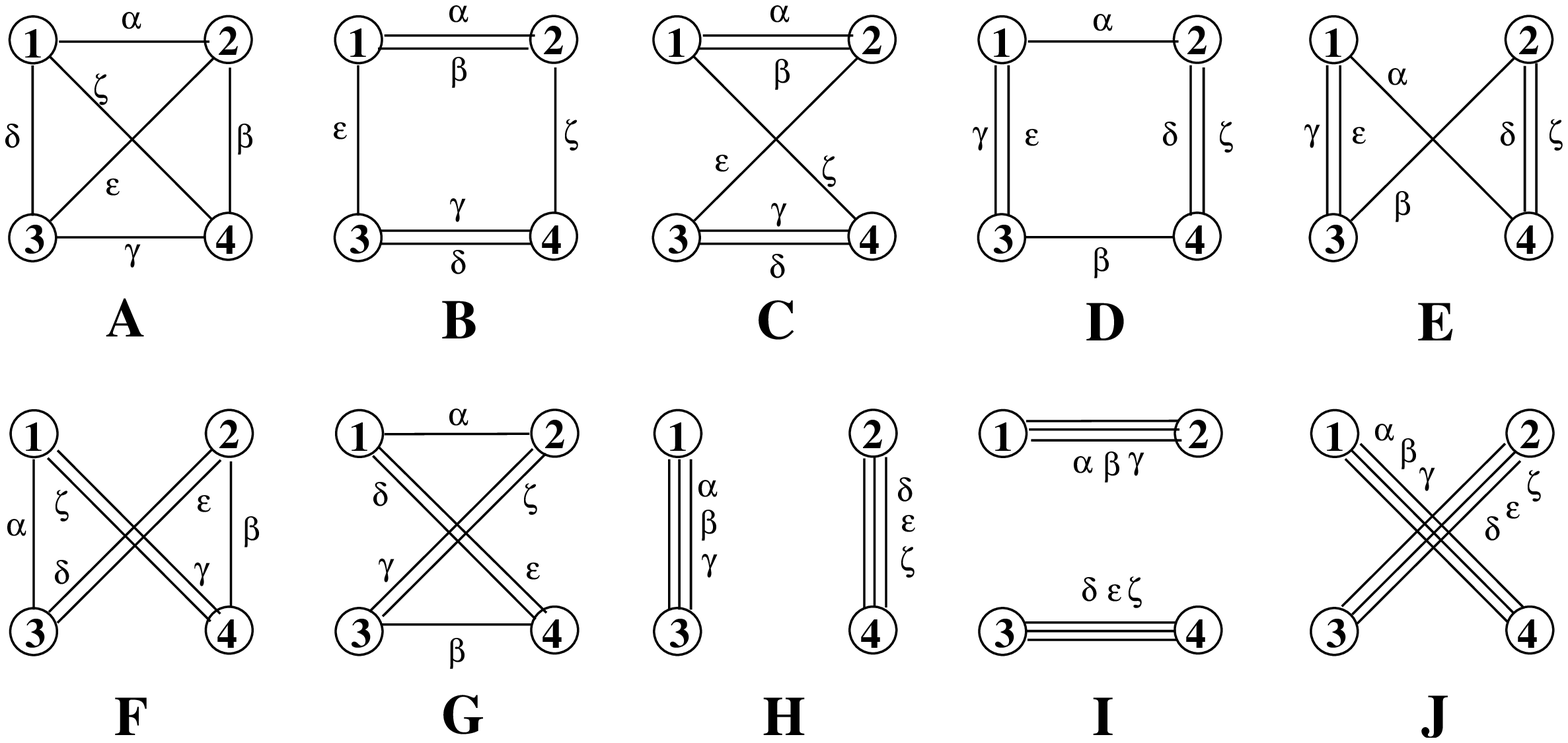}
}
}\caption{The 10 configurations arising from the randomization of the complete graph $\bf K^4$. In all configurations the strength $s$ of each vertex is preserved. However,  for $s=0,1,2$ one has $E_{>s}=6$ in {\bf A}, while $E_{>s}=4$ in {\bf B-G}, and  $E_{>s}=2$ in {\bf H-J}.}
\label{fig3a}
\end{figure}
The simplest example to describe our procedure is to start from a simple unweighted graph, where $w_m=1$. 
If the network is sparse the procedure is the analogous of the MSZ one. 
By contrast, dense unweighted graphs give rise to the creation of multiple edges. 
For instance, we illustrate the possible outcomes for a complete graph with $4$ vertices ($K^4$). 
Starting from the original configuration 
{\bf A} shown in Fig.\ref{fig3a}, the flipping of edges $\epsilon$ and $\zeta$ will produce 
either the configurations {\bf B} or {\bf D}. If different edges are selected, we find that starting from {\bf A} we can reach also the configurations {\bf C, E, F, G}. 
Similarly, starting from the configuration {\bf H}, 
we can flip any of the edges $\alpha, \beta, \gamma$ once with any of the edges $\delta, \epsilon, \zeta$. In one half of the cases we move to the configuration {\bf D}, in the 
other half to the configuration {\bf E}. Therefore $p_{HD}=p_{HE}=1/2$.
One can repeat this calculation for any other starting configuration keeping track of the various probabilities. 
In this way we can write the following transition matrix $T$:  
$$
\scriptsize{
T=
\left(
\begin{array}{cccccccccc}
0 & 2/5 & 2/5 & 2/5 & 2/5 & 2/5 & 2/5 & 0 & 0 & 0 \\
1/6 & 0 & 1/10 & 2/5 & 0 & 0 & 0 & 0 & 1/2 & 0 \\
1/6 & 1/10 & 0 & 0 & 0 & 0 & 2/5 & 0 & 1/2 & 0 \\
1/6 & 2/5 & 0 & 0 & 1/10 & 0 & 0 & 1/2 & 0 & 0 \\
1/6 & 0 & 0 & 1/10 & 0 & 2/5 & 0 & 1/2 & 0 & 0 \\
1/6 & 0 & 0 & 0 & 2/5 & 0 & 1/10 & 0 & 0 & 1/2 \\
1/6 & 0 & 2/5 & 0 & 0 & 1/10 & 0 & 0 & 0 & 1/2 \\
0 & 0 & 0 & 1/10 & 1/10 & 0 & 0 & 0 & 0 & 0 \\
0 & 1/10 & 1/10 & 0 & 0 & 0 & 0 & 0 & 0 & 0 \\
0 & 0 & 0 & 0 & 0 & 1/10 & 1/10 & 0 & 0 & 0 
\end{array}
\right). 
}
$$
Iterating $T$ yields the steady state matrix 
$T^\infty$, whose columns are all equal to 
each other and represent the vector of densities of the various configurations. 
We find $N({\bf A})=4/15, N({\bf B})=N({\bf C})=N({\bf D})=N({\bf E})
=N({\bf F})=N({\bf G})= 1/9, N({\bf H})=N({\bf I})=N({\bf J})=1/45$.
Note that in this case the strength is always an integer number, corresponding to multiple links ($w_m=1$). In particular, the strength of every vertex remains equal to the original value $3$. However, the $E_{>s}$ in our definition (\ref{eq_phiw}) counts all multiple edges created by the randomization as a single weighted edge.
Therefore, while in the original instance $\phi_w(s)=\phi(k)=1$ for every value of $s$ and $k$, across the randomized ensemble we obtain a $\phi^{NEW}_w(s) =11/15\simeq 0.73$ for $s=0,1,2$ smaller than the value $\phi^{MSZ}(k)=1$. 
This yields $\rho^w(s)=15/11\simeq 1.36$. In general, for complete graphs our procedure yields values of $\rho^w(s)$ systematically larger than the value $1$, correctly indicating the presence of the rich--club effect. 

\begin{figure}[t]
\centerline{
\resizebox{0.9\columnwidth}{!}{%
  \includegraphics{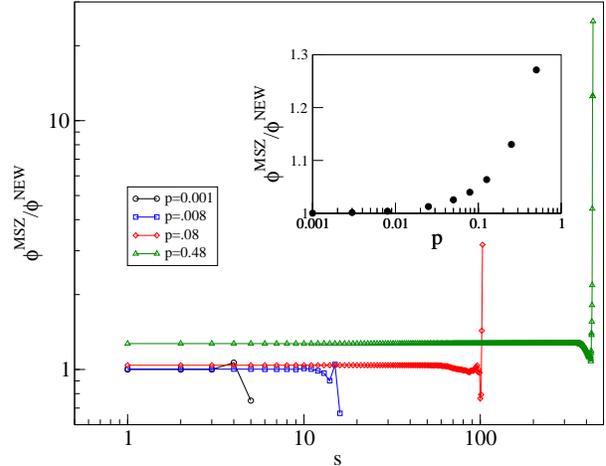}
}
}\caption{Rich--club analysis of random graphs. For sparse networks the ratio $\phi^{MSZ}/\phi_w$ is close to 1 and therefore our proposed randomization has the same desired properties as the MSZ procedure. When the density increases, $\phi_w$ becomes smaller than $\phi^{MSZ}$, meaning that $\rho^{MSZ}>\rho^w$. 
This is correct since in the limit of a complete graph the MSZ procedure can only yield $\phi^{MSZ}=\phi$, as no randomization is possible. However, this is the configuration for which one expects the maximum value of the rich--club coefficient.}
\label{fig4}
\end{figure}

In order to investigate the rich--club effect in the whole range between the extreme examples of sparse and fully connected networks, we  consider (Gilbert) Random Graphs \cite{Gilbert,ER60} 
corresponding to different values of the probability 
$p$ to draw an edge. We consider the range from $p=10^{-3}$ to $1/2$.
We study the difference between our method and the traditional unweighted one by considering the ratio of the two densities 
$\rho^w(s)/\rho^{MSZ}(k)$ computed with the present randomization and with the original MSZ algorithm. 
As previously noticed, when randomizing an originally unweighted graph the strength $s$ is given by the number of multiple edges that may be created in it. However, the original random graph has no such multiple edges, thus for the initial instance $k=s$ and $\phi_w(s)=\phi(k)$.
We can then write 
\begin{equation}
\frac {\rho^w(s)} {\rho^{MSZ}(s)}=\frac {\phi^{MSZ}(s)} {\phi^{NEW}(s)}
\end{equation}
which is independent of the original instance. For MSZ the argument $s$ always coincides with the original degree $k$ which is preserved, while this is not the case for our definitions.
The behaviour of the above ratio is shown in Fig.\ref{fig4} for various random graphs. In the main panel we show different plots corresponding to different choices of $p$. The trends are all characterized by a flat plateau whose height is larger than one. Therefore we find that our approach systematically yields $\phi^{MSZ} >\phi^{NEW}$, corresponding to $\rho^w<\rho^{MSZ}$. This effect becomes more pronounced as the density increases. This is clearly illustrated in the inset, where the height of the plateau shown in the main panel is now plotted as a function of $p$. As expected, for sparse graphs the ratio is close to one, since our generalized approach recovers the MSZ one. By contrast, for denser networks we correctly find that our algorithm explores a larger portion of configuration space, confirming the results discussed for the extreme example of $K^4$.

\section{Weighted graphs}
\begin{figure}[t]
\centerline{
\resizebox{0.9\columnwidth}{!}{%
  \includegraphics{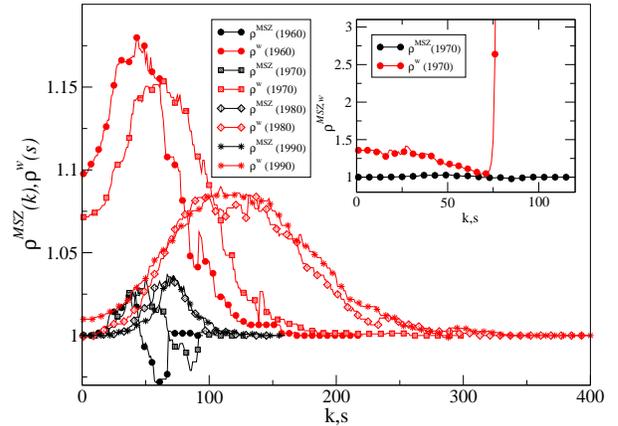}
}
}\caption{Main panel: our weighted rich--club coefficient $\rho^w$, compared with the unweighted one $\rho^{MSZ}$, for 4 different years of WTW when the network is considered weighted. 
Inset: same analysis on the unweighted version of the WTW for the year 1970. Other possible comparisons show similar discrepancies between the two procedures.}
\label{fig3}
\end{figure}

Finally, we illustrate our approach on inherently weighted networks. The example we consider is that of the World Trade Web (WTW) \cite{originalWTW,WTW2,chatta}. 
In this network the vertices are world countries, and a link represents a reciprocal trade relationship whose 
weight is given in millions of dollars.  
The data report trading exchanges between world countries for each year starting from 1948 to 2000  \cite{gleditsch}. 
The values of the weights span six orders of magnitude;  for this reason it is more convenient to redefine the weight as the logarithm of the original trade volume in dollars.
In the mail panel of Fig.\ref{fig3} we present the rich--club analysis on different snapshots of the weighted WTW.  
The first immediate result is that $\rho^{MSZ}(k)$ sometimes 
indicates anticorrelation between hubs. 
However, since in this region the number of configurations accessible by MSZ is 
really small, we interpret this feature as fluctuations in the  rich-club coefficient given by the poor statistics. 
In the case of $\rho^w(s)$ obtained with our algorithm, 
we instead identify a positive correlation between medium and large hubs as confirmed by 
a series of studies on the wealth and development of nations \cite{gleditsch,hidalgo}. This detects a significant correlation for countries with an intermediate strength value. Interestingly, this correlated region moves towards larger strength values as time proceeds from 1960 to 1990.
Furthermore we note that we can detect additional information, as compared with the MSZ algorithm, even if we ignore the information about the weights of the WTW. 
Indeed, if we discard the weights and simply consider the WTW as an unweighted network, then we can apply our strategy as already illustrated in the previous unweighted examples. For the year 1970, this is shown in the inset of Fig.\ref{fig3}. The original network has a connectance (number of actual edges divided by the number of possible ones) approximately equal to the very large value $0.5$. As usual, when using the traditional MSZ algorithm we obtain a constant value of $\rho^{MSZ}(k)$ close to one, mostly due to the small number of available random configurations. 
However, using our measure $\rho^{w}(s)$ we observe a decreasing trend towards $1$. This partially recovers the information on the positively correlated range for small and intermediate strength values. For larger values we find $\rho^{w}=1$, and no particular correlation can be extracted from the data. In this case, however, we can exclude that this feature is originated by the collapse of the available configuration space for randomized configurations.  

\section{Conclusions}
In conclusion, we have presented here a novel way to define a suitable 
randomization for dense and weighted networks. 
Various null models for weighted graphs have already been introduced  \cite{serrano2,colizza2,chatta,weighted,weightedconfiguration,SBP06}, depending on the properties one is interested in. 
Our emphasis here is on a problem arising when dense unweighted graphs are randomized while keeping the topological degrees fixed. In this case the collapse of the available configuration space has undesired effects on the rich club coefficient.  
We have therefore defined a procedure that solves this problem and at the same time provides a generalization of the rich--club property which is complementary to other recently proposed ones  \cite{serrano2,colizza2}. 
We have shown the advantage of our approach on unweighted graphs ranging from sparse to dense, and even complete. We have also shown its application to the real World Trade Web, both in its weighted and unweighted version. Our method outperforms the unweighted approach and  recovers significant information on the correlation structure of this network, which is in accordance with independent analyses of it.

G. C. acknowledges support from grant 2006PIV0001 (AGAUR of the Generalitat de Catalunya).

\end{document}